\documentclass[nofootinbib,twocolumn,a4paper,showpacs]{revtex4}
\usepackage{graphicx}
\usepackage{amsfonts}
\usepackage{amssymb}
\usepackage{amsmath}
\usepackage{latexsym}
\usepackage{subfloat}
\usepackage{dcolumn}

\begin{document}

\preprint{ICG preprint 04/xx}

\title{Ringing the Randall-Sundrum braneworld: metastable gravity wave bound states}

\author{Sanjeev S.~Seahra}
\affiliation{Institute of Cosmology \& Gravitation, University of
Portsmouth, Portsmouth, PO1 2EG, UK}

\setlength\arraycolsep{2pt}
\newcommand*{\di}{\partial}
\newcommand*{\V}{{\mathcal V}^{(k)}_d}
\newcommand*{\volume}{\sqrt{\sigma^{(k,d)}}}
\newcommand*{\OneTwo}{{(1,2)}}
\newcommand*{\onetwo}{{1,2}}
\newcommand*{\Lm}{{\mathcal L}_m}
\newcommand*{\stm}{{\textsc{stm}}}
\newcommand*{\Hm}{{\mathcal H}_m}
\newcommand*{\hatHm}{\hat{\mathcal H}_m}
\newcommand*{\Ldust}{{\mathcal L}_\mathrm{dust}}
\newcommand*{\maxsym}{{\mathbb S}_d^{(k)}}
\newcommand*{\sn}{{\mathrm{sn}}}
\newcommand*{\cn}{{\mathrm{cn}}}
\newcommand*{\nc}{{\mathrm{nc}}}
\newcommand*{\Jacobisc}{{\mathrm{sc}}}
\newcommand*{\ansatz}{{\emph{ansatz}}}
\newcommand*{\ds}[1]{ds^2_\text{\tiny{($#1$)}}}
\newcommand*{\kret}[1]{\mathfrak{K}_\text{\tiny{($#1$)}}}
\newcommand*{\ads}[1]{{AdS$_{#1}$}}
\newcommand*{\rhotot}{\rho_\text{tot}}
\newcommand*{\ptot}{p_\text{tot}}
\newcommand*{\rb}{r_\text{b}}
\newcommand*{\xb}{x_\text{b}}
\newcommand*{\kk}{\roarrow{k}}
\newcommand*{\Hank}[2]{\text{H}^{(#1)}_{#2}}

\date{September 12, 2005}

\begin{abstract}

In the Randall-Sundrum scenario, our universe is a 4-dimensional
`brane' living in a 5-dimensional bulk spacetime. By studying the
scattering of bulk gravity waves, we show that this brane rings
with a characteristic set of complex quasinormal frequencies, much
like a black hole.  To a bulk observer these modes are interpreted
as metastable gravity wave bound states, while a brane observer
views them as a discrete spectrum of decaying massive gravitons.
Potential implications of these scattering resonances are
discussed.

\end{abstract}

\pacs{11.25.Wx, 04.30.Nk \hfill hep-th/0501175}

\maketitle

\paragraph*{\textbf{Introduction}}

It is generally true that in many branches of physics, the best
way to learn about an object is to hit it with something and
analyze what comes back at you.  Scattering experiments can either
be real exercises done in the laboratory, such as in particle or
condensed matter physics, or they can be of the virtual type,
where the behaviour of radiation striking a target is simulated on
a computer.  The latter technique has proved to be an invaluable
tool in the analysis of the fundamental structure of black holes,
which are a class of targets not readily available in the lab.
Numerical calculations of how gravity wave pulses bounce off black
holes have revealed that they possess discrete quasinormal modes
(QNMs) of vibration \cite{Vish}, which encode both important
observational features for gravity wave detectors and valuable
clues as to the canonical quantization of general relativity (GR)
\cite{Dreyer}.

The success achieved for black holes prompts us to look for other
virtual gravity wave scattering targets.  A prime candidate comes
from the Randall-Sundrum (RS) braneworld scenario \cite{RS}. This
5-dimensional phenomenological model is motivated by
non-perturbative ideas from string theory, which postulate that we
are living on a hypersurface (or `brane') embedded within a
higher-dimensional spacetime.  One of the two variants of the
model involves one flat brane and an infinite fifth dimension.
Usually, models with non-compact extra dimensions are not viable
because particles and fields exhibit higher-dimensional behaviour
that is inconsistent with what we see in the world around us.
However, the RS model escapes this fate via several key
assumptions. First, standard model fields are confined on the
brane, and hence behave in an entirely 4-dimensional manner.
Second, although gravity is allowed to propagate in the bulk, at
low energies it is effectively trapped on the brane. This is
achieved by a negative cosmological constant in the bulk and
enforcing the $\mathbb{Z}_2$ reflection symmetry across the brane.
The result is that the graviton has a continuum of modes that have
an effective 4-dimensional mass, and a single localized and
normalizable massless mode. The latter `zero' mode reproduces GR
at low energies, while the former `Kaluza-Klein' modes predict
high-energy corrections to it.

In this article, we consider the problem of scattering gravity
waves off the brane in the RS scenario.%
\footnote{We note that wave propagation (gravitational and
otherwise) around 4-dimensional domain walls \cite{4d} and
supergravity `D-brane' solutions
\cite{D-brane} has been considered by several authors.}%
We will demonstrate that just like black holes, the brane
possesses a discrete set of QNMs that appear as scattering
resonances; i.e., complex poles of the scattering matrix.  Since
scattering resonances are essentially metastable bound states,
this result has an important phenomenological implication for the
RS one-brane scenario: namely, within the continuum of massive
Kaluza-Klein graviton states, there are a discrete set of resonant
modes that remain localized near the brane with definite
lifetimes.  After confirming the existence of these modes both
analytically and numerically, we discuss how they might impact on
various aspects of the braneworld scenario.

\paragraph*{\textbf{Gravity wave master equation}}

The RS scenario consists of an AdS$_5$ spacetime with a geometric
defect (brane) at $y=0$.  The 5-metric is
\begin{equation}
    ds^2 = a^2(y) \eta_{\mu\nu} d\mathsf{x}^\mu d\mathsf{x}^\nu + dy^2, \quad a(y) =
    e^{-|y|/\ell}.
\end{equation}
As is well known, in the RS gauge perturbations of the 5-metric
are written as
\begin{equation}
    g_{ab} \rightarrow g_{ab} + \di_a \mathsf{x}^\mu \di_b \mathsf{x}^\nu
    h_{\mu\nu}, \quad \eta^{\mu\nu} h_{\mu\nu} = 0 = \di^\mu
    h_{\mu\nu},
\end{equation}
subject to the boundary condition $\di_y h_{\mu\nu} = -2\ell^{-1}
h_{\mu\nu}$ at $y =0$.  The simplest class of perturbations are
the tensor modes, with harmonics ($i,j=1,2,3$):
\begin{equation}
    \roarrow{\nabla}^2 \mathbb{T}_{ij}^{(\kk)} = -k^2 \mathbb{T}^{(\kk)}_{ij}, \quad
    \di^i \mathbb{T}^{(\kk)}_{ij} = 0 = \mathrm{Tr} \,
    \mathbb{T}^{(\kk)}.
\end{equation}
Then, we express the metric perturbation as
\begin{equation}\label{expansion}
    \delta h_{ij} =
    \int \frac{d^3\kk}{(2\pi)^3} e^{-|y|/2\ell} \Phi_k(t,y) \mathbb{T}^{(\kk)}_{ij}, \quad \delta h_{0\mu} = 0.
\end{equation}
Putting this \emph{ansatz} into the linearized Einstein equations
$\delta G_{ab} = (6/\ell^2)\delta g_{ab}$ yields that the $\Phi_k$
expansion coefficients satisfy a master wave equation for $y > 0$:
\begin{subequations}
\begin{eqnarray}\label{wave equation}
    0 & = & \left[ \di_\tau^2 - \di_x^2
    + V_k(x) \right] \Phi_k, \,\, V_k(x) = \kappa^2 + \tfrac{15}{4}
    x^{-2}, \\
    \label{boundary condition}
    0 & = & \left( \di_x + \tfrac{3}{2} \right) \Phi_k \text{ at } x = 1.
\end{eqnarray}
\end{subequations}
Here, we have defined dimensionless coordinates $\tau = t/\ell$
and $x = e^{y/\ell} > 1$, as well as the dimensionless wavenumber
$\kappa = k\ell$.  Note that the boundary condition (\ref{boundary
condition}) and the $x < 1$ region can be incorporated into the
potential via a delta-function and the imposition of mirror
symmetry:
\begin{equation}\label{potential}
    V_k(x) = \kappa^2 + \tfrac{15}{4}\hat{x}^{-2} - 3 \, \delta(\hat{x}-1), \quad \hat{x} \equiv
    |x-1|+1.
\end{equation}
So, the problem of perturbing the RS one-brane scenario is
equivalent to the problem of wave propagation in a one-dimensional
potential.

\paragraph*{\textbf{Quasinormal resonances}}

\begin{figure}
\includegraphics[width=0.9\columnwidth]{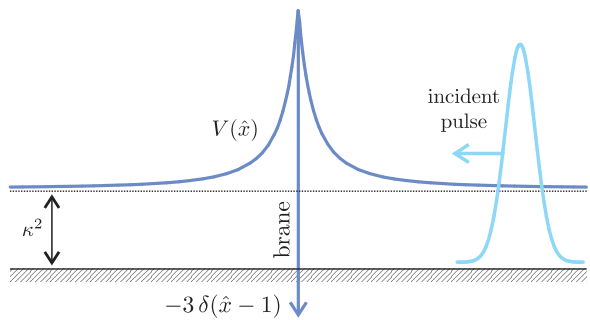}
\caption{A schematic of the brane potential and incident gravity
wave pulse in our scattering problem}\label{fig:potential}
\end{figure}

The situation we are interested in is shown in Figure
\ref{fig:potential}. We regard the spatial wavenumber $\kappa$ of
the perturbation as given, and we consider a compact pulse of
right-moving gravitational radiation centered about $x_0 > 1$ at
$t=0$.  This pulse will of course strike the brane and scatter,
and we are interested in the late-time features of the reflected
waveform. Now, since the potential governing gravity waves
(\ref{potential}) is centered about an attractive delta-function,
we can reasonably expect some of the incident wave to become
trapped and remain localized near the brane forever, while the
rest of energy will bounce off and propagate to infinity.

However, all the reflected energy need not be \emph{promptly}
reflected away from the brane.  A well-known phenomenon in
spectral theory is that of a \emph{scattering resonance}
\cite{scattering ref}. This occurs when certain frequency
components of the initial pulse become temporarily trapped within
the potential. One can think of these modes as metastable bound
states: they are initially localized, but they eventually leak out
to infinity.  In black hole perturbation theory, these metastable
states are precisely the QNMs. In that problem, the master
equation has the form of (\ref{wave equation}) with a potential
consisting of a smooth barrier centered about the photon-sphere.
The late time behaviour of a pulse after bouncing off this barrier
is dominated by QNMs with $\exp(i\omega_n\tau)$ time-dependence,
where $\{\omega_n\}$ is a discrete set of \emph{complex}
frequencies with $\text{Im} \,\, \omega_n > 0$; i.e., quasinormal
oscillations are damped in time.

\begin{figure}
\includegraphics[width=0.9\columnwidth]{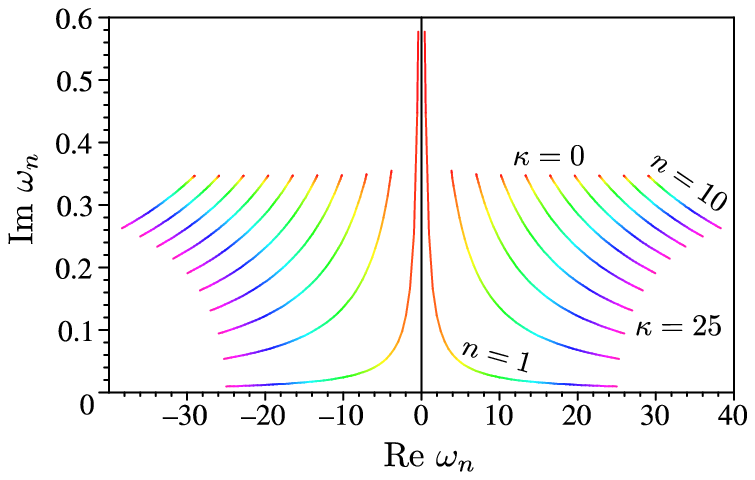}
\caption{The first twenty quasinormal frequencies of the brane as
the spatial wavenumber $\kappa$ is varied from 0 (red) to 25
(purple).%
Those in the right half of the complex plane correspond
to $\omega_n^{(1)}$, while those in the left half are
$\omega_n^{(2)}$
}\label{fig:QNMs}
\end{figure}

The question is now: does the gravitational wave potential in the
RS wave equation (\ref{wave equation}) support QNMs? To answer, we
first assume $\Phi_k=e^{i\omega\tau} \phi_k(x)$, which yields the
Schr\"odinger-like equation: $\omega^2 \phi_k = -\phi_k'' + V_k(x)
\phi_k$. If $\omega^2 \ne \kappa^2$, this admits an exact solution
that satisfies the boundary condition (\ref{boundary condition})
at the brane:
\begin{equation}
    \phi_k(x) = \sqrt{\hat{x}} \left[ \Hank{1}{1}(\mu) \Hank{2}{2}(\mu \hat{x}) -
    \Hank{2}{1}(\mu) \Hank{1}{2}(\mu \hat{x}) \right].
\end{equation}
Here, $\Hank{1}{\nu}$ and $\Hank{2}{\nu}$ are Hankel functions of
the first and second kind respectively, and $\nu$ is their order.
The dimensionless $\mu$ parameter is defined via the dispersion
relation
\begin{equation}
    \mu^2 = \omega^2 - \kappa^2, \quad |\text{Arg}(\mu)|<\pi.
\end{equation}
Conventionally, $\mu$ is assumed to be real and positive, and is
related to the conventional Kaluza-Klein mass of the fluctuation
via $\mu = m\ell$.  However, the key to finding a resonant mode is
to let $\mu$ range over the complex plane.  Consider the
asymptotic expansion of the Hankel functions:
\begin{equation}
    \phi_k(x) \propto \Hank{1}{1}(\mu) e^{-i\mu x} -
    \Hank{2}{1}(\mu) e^{+i\mu x} \text{ as } x \rightarrow \infty.
\end{equation}
Depending on the sign of $\text{Re} \,\, \mu$ relative to the sign
of $\text{Re}\,\,\omega$, each term corresponds to either a left
or right moving wave at infinity.  The next step in the QNM
programme is to demand that the asymptotic form of $\phi_k$ is
`purely outgoing' \cite{Vish}, which in this context means that we
need to set the coefficient of the left-moving wave to
zero.%
\footnote{The fact that QNMs are defined by a boundary
condition `at infinity' tells us that they are not relevant to
two-brane scenarios where the extra dimension is effectively
finite. Generally speaking, in order for a system to exhibit
metastable resonances there must be some type of mechanism to
remove energy from the system, which only really happens in the
one-brane case.}%
Formally, we are searching for a complex pole of
the scattering matrix $S_k(\omega)$, which is defined as the ratio
of the amplitudes of outgoing and ingoing radiation at infinity.
Clearly, for a pole we need $\mu$ to be a zero of either
$\Hank{1}{1}$ or $\Hank{2}{1}$. For reference, we quote the first
four zeros of $\Hank{1}{1}$:
\begin{displaymath}
\begin{array}{cD{.}{.}{12}ccD{.}{.}{12}}
\mu_1 = & -0.419 -0.577\,i,&\,\, &
\mu_2 = & -3.832 -0.355\,i, \\
\mu_3 = & -7.016 -0.350\,i,& & \mu_4 = & -10.174-0.348\,i.
\end{array}
\end{displaymath}
The corresponding zeros of $\Hank{2}{1}$ are simply the complex
conjugates $\bar{\mu}_n$, and it is easy to confirm that all zeros
of both Hankel functions have negative real parts.  Therefore, if
$\text{Re}\,\,\omega > 0$ we must set $\Hank{1}{1}(\mu)=0$ in
order to have a purely outgoing wave; that is, we choose $\mu \in
\{\mu_n\}$. Conversely, if $\text{Re}\,\,\omega < 0$ we need
$\Hank{2}{1}(\mu) = 0$, or in other words $\mu \in
\{\bar{\mu}_n\}$. From this, it follows that there are two
families of complementary quasinormal frequencies:
\begin{subequations}
\begin{eqnarray}
    \omega^{(1)}_n(\kappa) & = & +\sqrt{\mu_n^2+\kappa^2}, \quad
    \Hank{1}{1}(\mu_n)=0, \\
    \omega^{(2)}_n(\kappa) & = & -\sqrt{\bar{\mu}_n^2+\kappa^2}, \quad
    \Hank{2}{1}(\bar{\mu}_n)=0.
\end{eqnarray}
\end{subequations}
The variation of the first twenty QNMs with $\kappa$ is depicted
in Figure \ref{fig:QNMs}.  Note that in order to evaluate the
square root, we need to take the same branch cut as that of the
Hankel function; i.e., the negative real axis.  In particular,
this means $\sqrt{\mu_n^2} = -\mu_n$.

Before moving on, we should mention that the Randall-Sundrum model
is not the only 5-dimensional scenario that exhibits resonant
behaviour.  The Dvali-Gabadadze-Porrati (DGP) model features a
Minkowski bulk surrounding a brane with `induced gravity'
\cite{DGP}. There is no zero-mode in this model, but the massive
modes sum together in such a way to approximately recover Newton's
law on \emph{small} scales.  This superposition of massive modes
is interpreted as a metastable 4-dimensional graviton that decays
into the bulk after a finite streaming length along the brane.  To
the best of our knowledge, the QNM spectrum of the DGP model has
not been specifically investigated, although for certain branch
cut choice one can find DGP resonances with $\text{Im}\,\omega <
0$; i.e., unstable fluctuations \cite{DGP stability}.  An
interesting and perhaps fruitful exercise would be to repeat the
current calculation for that model.

\paragraph*{\textbf{Numerical results}}

\begin{figure}
\includegraphics[width=0.9\columnwidth]{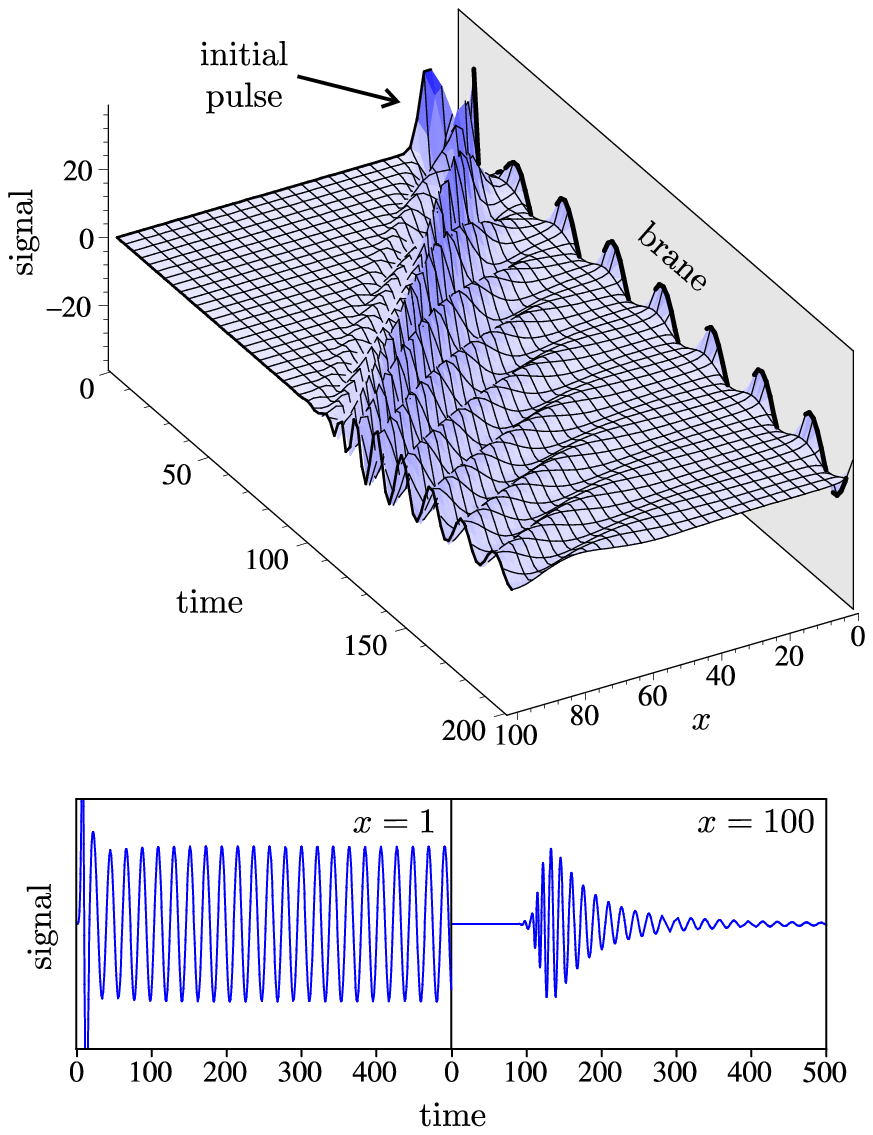}
\caption{Numeric waveform for $\kappa = \pi/10$ and an incident
Gaussian pulse centered at $x=10$ with a FWHM of
5}\label{fig:profile}
\end{figure}

So far, we have concentrated on modes that are outgoing at spatial
infinity.  But the RS potential also supports a normalizable bound
state:
\begin{equation}
    \Phi^{(0)}_k(\tau,x) = e^{i\omega\tau} \hat{x}^{-3/2}, \quad
    \omega = \pm \kappa.
\end{equation}
The existence of this so-called `zero-mode' is what gives the RS
model its main appeal, since it gives rise to GR at low energies.
How does the zero mode manifest itself in our scattering problem?
To answer, we first look at a numeric simulation of a Gaussian
pulse incident on the brane in Figure \ref{fig:profile}. This is
obtained by numerically integrating (\ref{wave equation}) for $x >
1$ and with the boundary condition (\ref{boundary condition}). Our
assumed initial data is at $x=10$ and has a full-width at half
maximum (FWHM) of 5, and we take $\kappa = \pi/10$. We see that
after the pulse hits the brane, part of the energy is reflected
and propagates off to infinity while the other part remains
trapped on the brane at $x=1$. Indeed, a close examination of the
brane perturbation shows that after the initial pulse, the
waveform is an undamped sinusoid with frequency $\pi/10$.  On the
other hand, the waveform at $x=100$ shows that the reflected pulse
damps to zero as $\tau \rightarrow \infty$. In other words, the
incident pulse excites a perpetual and spatially localized
perturbation on the brane; i.e., the zero-mode.

The one thing missing from the situation depicted in Figure
\ref{fig:profile} is any hint of resonant phenomena.  The Gaussian
pulse immediately excites the zero mode when it strikes the brane,
which suggests any quasinormal resonances are sub-dominant to the
undamped bound state.  Some experimentation reveals that this is
true for many different types of Gaussians and values of $\kappa$.
The reason is simple: in Fourier space a Gaussian pulse always has
a significant low frequency component that will efficiently excite
the zero mode.  In order to see the QNM influence directly, we
should really consider coherent plane wave initial data with a
frequency high enough to preferentially excite the first QNM.  But
we cannot model an infinite plane wave numerically, so we instead
use a finite wavetrain, which is essentially a segment of the
left-moving wave $\sin(\omega_0 \tau + \mu_0 x)$, with $\omega_0^2
= \mu_0^2 + \kappa^2$.  If such a wavetrain is $N$ cycles long,
its spectral width is on the order of $\Delta\omega = \omega_0/N$.
It is also useful to look at the reflected wave from an off-brane
vantage point, because the influence of the zero mode is strongest
at $x=1$.

\begin{figure}
\includegraphics{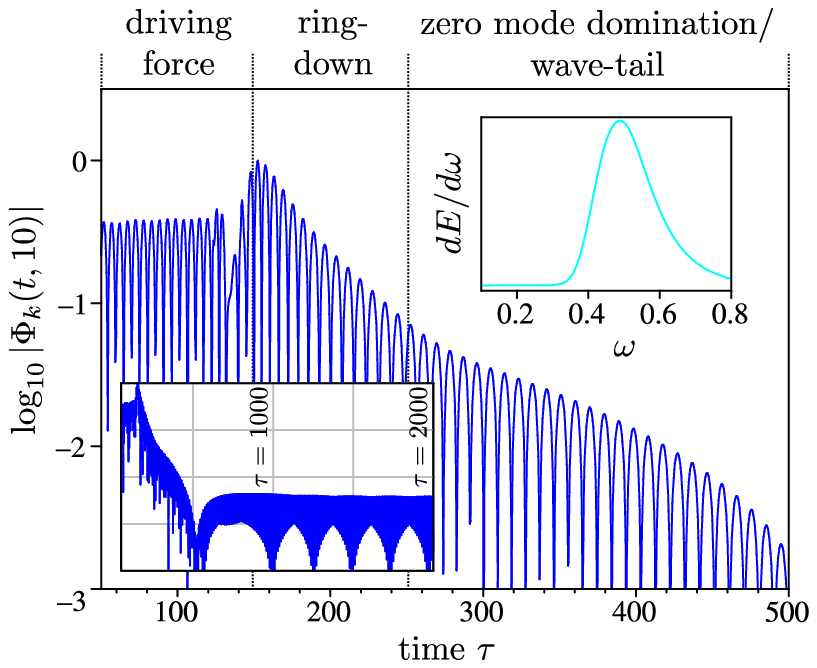}
\caption{Numeric waveform evaluated at $x = 10$ for an incident
wavetrain with $\omega_0 = 0.6$ and 10 wavelengths long.  We take
$\kappa = \pi/10$.  The upper inset shows the power spectrum of
the waveform during the ringdown phase, while the lower inset
shows its late time behaviour}\label{fig:wavetrain}
\end{figure}

In Figure \ref{fig:wavetrain}, we show the perturbation evaluated
at $x = 10$ for this type of initial data with $\kappa = \pi/10$
and $\omega_0 = 0.6$.  The lefthand edge of the wavetrain is
initially at $x = 10$, and it is 10 wavelengths long. For this
choice of $\kappa$, the first quasinormal mode is at
$\omega_1^{(1)} = 0.463 + 0.523\,i$.  In this case, the
gravitational waveform is analogous to the response of a damped
mechanical oscillator subject to a driving force.  The natural
frequency of the system is that of the zero mode $\omega = \pi/10
\sim 0.314$, so the wavetrain represents a high frequency forcing
term. As the wavetrain reflects off the brane, the perturbation is
obliged to oscillate at $\omega = 0.6$. By $t \sim 150$, the
entire reflected wavetrain has travelled past our vantage point,
and the driving force is effectively `switched off.'  But the
waveform does not immediately revert to zero mode vibrations,
there is instead a `ringdown phase' where the brane radiates away
the high frequency components of the driving force. The upper
inset of Figure \ref{fig:wavetrain} shows the amplitude squared of
the waveforms's Fast Fourier Transform (FFT) during this ringdown.
The power spectrum is peaked at the real part of the first QNM at
0.463, which is fairly clear evidence of quasinormal ringing of
the brane. (Note that the damping of a QNM means that its power
spectrum has a significant width.)  After the ringdown phase, the
FFT (not shown) is sharply peaked about $\pi/10$, suggesting
strong domination of the zero mode.  In the standard lexicon of
black hole perturbation theory, the part of the waveform that
follows quasinormal ringing is known as the `wave-tail.'  So in
the RS braneworld, the wave tail essentially consists of zero mode
oscillations.  The lower inset in Figure \ref{fig:wavetrain} shows
that the amplitude of the zero mode signal becomes constant only
for late times.  Note that a slow variation in the amplitude of
oscillatory wave-tails is a familiar feature of potentials that
are non-vanishing at infinity \cite{wavetail}.  In the future, we
will report on the Randall-Sundrum wave-tail in detail.

\paragraph*{\textbf{Implications}}

As mentioned above, the principle implication of the existence of
QNMs in the RS one-brane scenario is the prediction of a discrete
set of metastable massive modes within the continuous Kaluza-Klein
tower.%
\footnote{Alternatively, one could view the resonant modes as
spin-2 fields obeying a complex-valued dispersion relation; i.e.,
travelling in an absorptive medium.  Details of such an
interpretation can be found in Ref.~\cite{Seahra:2005iq}.}%
If $\ell = 0.1$ mm, which is an upper value set by experimental
limits on deviations from Newton's law \cite{Hoyle:2004cw}, the
lowest order resonance has mass $8.386 \times 10^{-4}$ eV and a
half life of $4.002 \times 10^{-11}$ seconds; i.e., it is light
and short-lived. Ideally, one would look for such resonances in
particle accelerators, but in the RS one-brane case the coupling
between matter and the Kaluza-Klein states is quite small since
the 5-dimensional Planck mass is $\gtrsim 10^5$ TeV.  This scale
suggests that the best place to look for QNMs is in the
high-energy regime associated with the early universe; in
particular, one can expect the stochastic gravity wave background
\cite{Kazuya} to carry potentially observable echoes of metastable
masses. Of course, the expansion of the brane universe at that
epoch will certainly change the spectrum of quasinormal
resonances.   However, it is reasonable to assume that very
high-frequency QNMs will `see' a moving brane to be stationary,
and hence our results from the static case will be applicable.

It is interesting, and perhaps instructive, to see what
contribution the quasinormal modes make to the effective
braneworld Einstein equations.  Using the Shiromizu-Maeda-Sasaki
formalism \cite{Shiromizu:1999wj}, we see that QNMs induce `dark
radiation' on the brane---essentially the pullback of the
5-dimensional Weyl tensor onto the 4-dimensional spacetime. The
QNM contribution to the stress energy of this effective matter
field is
\begin{equation}\label{Weyl}
    {\mathcal{E}}^\text{\textsc{QNM}}_{\mu\nu}(\kk) = \sum_{n} A_n
    \mu_n e^{i\omega_n \tau} \mathbb{T}_{\mu\nu}^{(\kk)},
\end{equation}
where $A_n$ are constants describing the degree to which mode is
excited.  Note the explicit dependance on the complex mass
$\mu_n$; the massless zero-mode does not enter into the dark
radiation. Hence, the quasinormal resonances can be interpreted as
a discrete set of brane matter fields prone to decay by tunnelling
into the bulk. Returning to the cosmological case, this implies
that the principal high frequency bulk effects on early universe
cosmological perturbations are via `matter' contributions of the
form (\ref{Weyl}).  The quantitative effects of these fields on
the cosmic microwave background and large scale structure are yet
to be calculated.

Finally, Eq.~(\ref{Weyl}) hints at a possible holographic
interpretation of these resonances.  According to the AdS/CFT
conjecture, the Randall-Sundrum model is dual to a large-$N$
conformal field theory (CFT) with cutoff living on the brane
\cite{adscft}.  Specifically, the retarded bulk graviton Green's
function is identified with the generating functional of the CFT.
Now, the metastable resonances found here are complex poles of the
Green's function, and hence will play an important role in the
boundary theory.  In particular, note that $\mathcal{E}_{\mu\nu}$
in the gravitational theory is precisely the stress-energy tensor
of the CFT, which suggests that the gravitational QNMs describe
how the boundary theory relaxes to equilibrium after some initial
disturbance, and also give the timescale for such a relaxation.

\paragraph*{\textbf{Summary}} Just like a black hole, the RS
braneworld `rings' with a discrete set of complex frequencies.
These metastable bound states, which look like decaying massive
gravitons to 4-dimensional observers, should play an important
role in any brane phenomena involving bulk degrees of freedom. An
interesting question is: how do QNMs manifest themselves in
generalizations of the RS scenario? We have recently reported on
the resonant behaviour of a spherical Einstein static brane
encircling a Schwarzschild-AdS$_5$ black hole \cite{ES}, but the
key challenge is characterizing the quasinormal excitations of a
moving cosmological brane. We hope to investigate this in the
future.

\paragraph*{\textbf {Acknowledgements}}

I would like to thank C.~Clarkson, K.~Koyama, R.~Maartens,
A.~Mennim, F.~Piazza, and D.~Wands for discussions and NSERC for
financial support.

\end{document}